\documentclass[12 pt, amsfonts, amssymb,color]{article}

\usepackage{amsmath,amssymb,amsfonts,latexsym,float,graphics,epsfig}

\usepackage{subcaption}
\def\R{\mathbb{R}}

\begin{document}

\renewcommand\theequation{\arabic{section}.\arabic{equation}}
\catcode`@=11 \@addtoreset{equation}{section}
\newtheorem{axiom}{Definition}[section]
\newtheorem{theorem}{Theorem}[section]
\newtheorem{axiom2}{Example}[section]
\newtheorem{lem}{Lemma}[section]
\newtheorem{prop}{Proposition}[section]
\newtheorem{cor}{Corollary}[section]
\newcommand{\be}{\begin{equation}}
\newcommand{\ee}{\end{equation}}

\newcommand{\equal}{\!\!\!&=&\!\!\!}
\newcommand{\rd}{\partial}
\newcommand{\g}{\hat {\cal G}}
\newcommand{\bo}{\bigodot}
\newcommand{\res}{\mathop{\mbox{\rm res}}}
\newcommand{\diag}{\mathop{\mbox{\rm diag}}}
\newcommand{\Tr}{\mathop{\mbox{\rm Tr}}}
\newcommand{\const}{\mbox{\rm const.}\;}
\newcommand{\cA}{{\cal A}}
\newcommand{\bA}{{\bf A}}
\newcommand{\Abar}{{\bar{A}}}
\newcommand{\cAbar}{{\bar{\cA}}}
\newcommand{\bAbar}{{\bar{\bA}}}
\newcommand{\cB}{{\cal B}}
\newcommand{\bB}{{\bf B}}
\newcommand{\Bbar}{{\bar{B}}}
\newcommand{\cBbar}{{\bar{\cB}}}
\newcommand{\bBbar}{{\bar{\bB}}}
\newcommand{\bC}{{\bf C}}
\newcommand{\cbar}{{\bar{c}}}
\newcommand{\Cbar}{{\bar{C}}}
\newcommand{\Hbar}{{\bar{H}}}
\newcommand{\cL}{{\cal L}}
\newcommand{\bL}{{\bf L}}
\newcommand{\Lbar}{{\bar{L}}}
\newcommand{\cLbar}{{\bar{\cL}}}
\newcommand{\bLbar}{{\bar{\bL}}}
\newcommand{\cM}{{\cal M}}
\newcommand{\bM}{{\bf M}}
\newcommand{\Mbar}{{\bar{M}}}
\newcommand{\cMbar}{{\bar{\cM}}}
\newcommand{\bMbar}{{\bar{\bM}}}
\newcommand{\cP}{{\cal P}}
\newcommand{\cQ}{{\cal Q}}
\newcommand{\bU}{{\bf U}}
\newcommand{\bR}{{\bf R}}
\newcommand{\cW}{{\cal W}}
\newcommand{\bW}{{\bf W}}
\newcommand{\bZ}{{\bf Z}}
\newcommand{\Wbar}{{\bar{W}}}
\newcommand{\Xbar}{{\bar{X}}}
\newcommand{\cWbar}{{\bar{\cW}}}
\newcommand{\bWbar}{{\bar{\bW}}}
\newcommand{\abar}{{\bar{a}}}
\newcommand{\nbar}{{\bar{n}}}
\newcommand{\pbar}{{\bar{p}}}
\newcommand{\tbar}{{\bar{t}}}
\newcommand{\ubar}{{\bar{u}}}
\newcommand{\utilde}{\tilde{u}}
\newcommand{\vbar}{{\bar{v}}}
\newcommand{\wbar}{{\bar{w}}}
\newcommand{\phibar}{{\bar{\phi}}}
\newcommand{\Psibar}{{\bar{\Psi}}}
\newcommand{\bLambda}{{\bf \Lambda}}
\newcommand{\bDelta}{{\bf \Delta}}
\newcommand{\p}{\partial}
\newcommand{\om}{{\Omega \cal G}}
\newcommand{\ID}{{\mathbb{D}}}
\newcommand{\pr}{{\prime}}
\newcommand{\prr}{{\prime\prime}}
\newcommand{\prrr}{{\prime\prime\prime}}
\title{ Monotonicity of the period function of the Li\'{e}nard equation of  second kind \\}
\author{A Ghose-Choudhury\footnote{E-mail aghosechoudhury@gmail.com}\\
Department of Physics, Surendranath  College,\\ 24/2 Mahatma
Gandhi Road, Calcutta 700009, India\\
\and
Partha Guha\footnote{E-mail: partha@bose.res.in}\\
SN Bose National Centre for Basic Sciences \\
JD Block, Sector III, Salt Lake \\ Kolkata 700098,  India \\
}

\date{ }

 \maketitle

\smallskip

\smallskip

\begin{abstract}
\textit{This paper is concerned with the monotonicity of the period function
for closed orbits of systems of the Li\'{e}nard II type equation given by
 $\ddot{x} + f(x)\dot{x}^{2} + g(x) = 0$.
We generalize  Chicone's result regarding
 the monotonicity of the period function to planar Hamiltonian vector fields in the presence of a position
dependent mass. Sufficient conditions  are also  given  for the isochronicity of the potential in case of such a system. }
\end{abstract}

\smallskip

\paragraph{Mathematics Classification (2010)}:34C14, 34C20.

\smallskip


\section{Introduction}
Investigations of the period function of second-order ordinary differential equations (ODEs) are of considerable interest as they shed light on the nature of the evolution of the system under consideration as also on the existence of periodic orbits and whether such orbits display isochronous behaviour. A Li\'{e}nard equation of the second kind is given by
\be\label{L2} \ddot{x} +f(x) \dot{x}^2 +g(x)=0,\ee
and arises naturally from Newtonian considerations when the mass of a particle is position dependent. The standard Li\'{e}nard equation
 $$\ddot{x} +f(x) \dot{x}+g(x)=0,$$
 on the other hand,  is related to damping and may be viewed as a generalization of the equation of a damped oscillator when the damping coefficient  is position dependent. The Li\'{e}nard equation of the second kind admits a Lagrangian interpretation with
 \be L=\frac{1}{2}\mu(x)\dot{x}^2-V(x).\ee
 The corresponding classical  Hamiltonian is
 \be\label{H1} H=\frac{p_x^2}{2\mu(x)}+V(x),\ee and leads to the Hamilton's equations:
 \be\label{H2} \dot{x}=\frac{p_x}{\mu(x)}, \;\;\;\dot{p_x}=\frac{\mu^\prime(x)}{2\mu^2(x)}p_x^2-V^\prime(x).\ee
 Eqn (\ref{H2}) are equivalent to the second-order ODE
 \be\label{L3} \ddot{x}+\frac{\mu^\prime(x)}{2\mu(x)}\dot{x}^2+\frac{V^\prime(x)}{\mu(x)}=0\ee and may be compared with (\ref{L2}) from which it is easy to deduce that
 \be\label{1.5a} \mu(x)=\exp(2\int f(x) dx)\;\;\;\mbox{and}\;\;\;V(x)=\int \mu(x) g(x) dx.\ee Furthermore it will be assumed that $V(x)$ is a smooth function having a nondegenerate relative minimum at the origin or in other words $V(0)=V^\prime(0)=0$ with $V^{\prime\prime}(0)>0$.
  One can verify that $dH/dt$ vanishes and hence $H$ is a constant of motion.
 The vector field associated with the Hamiltonian $H$, \textit{viz}
 \be X_H=\frac{p_x}{\mu(x)}\frac{\partial}{\partial x}+\left(\frac{\mu^\prime(x)}{2\mu^2(x)}p_x^2-V^\prime(x)\right)
 \frac{\partial}{\partial p_x},\ee is symmetric with respect to the $x$-axis and it is assumed that the phase portrait has a centre at the origin which is surrounded by periodic orbits each of which corresponds with the level curves of energy $E\in (0, E^*)$. The period function for a closed orbit is defined as $T:(0, E^*)\rightarrow \R$ and represents the minimum time required to traverse a periodic orbit. \\

 A substantial body of work has been devoted to understanding
the behavior of the period function in case of planar differential systems. The case of planar quadratic isochronous system were completely classified by Loud \cite{Loud}. Chow and Sanders [CsS] have shown that
systems of the form $\ddot{x} + g(x) = 0$ with $g(x)$ quadratic, necessarily have
monotone period function. It is worth noting  that  many different methods of analysis have been
applied to study the period function by different authors such as Chicone \cite{C1,C2}
Chicone and Dumortier \cite{CD1,CD2}, Chow and Sanders \cite{CS},Chow and Wang \cite{CW, W1},
Francoise \cite{F1}, Mardesic et al. \cite{Mardesic1,Mardesic2}, Sabatini \cite{S1}, 
Schaaf \cite{Sr1, Sr2}, Waldvogel \cite{WJ1, WJ2}, Zhao \cite{Zhao}.\\

A constructive characterizations
of local isochronous centers was proposed by Urabe \cite{Urabe}, in terms of
a differential equation that must be satisfied by the potential $V$, and that contains an arbitrary odd
function. Urabe's method has been cited, developed and re-examined by many authors, recently
Chouikha \cite{Chouikha2,Strelcyn} put forward an alternative notion of isochronicity condition.\\

\textbf{Result and motivation:}\\

 The monotonicity of the period function for systems of the form $\dot{x}=-y,\dot{y}=V^\prime(x)$ i.e., when $\mu(x)=1$ was studied by Chicone in \cite{C1}. The introduction of the mass function $\mu(x)$ leads to a generalization of Chicone's earlier result. Our main result may be stated as follows. Let $K(E)=\{x\in \R: V(x)\le E\}$ and assume $\mu(x)>0$ for all $x\in K(E)$. Define a function
 $$N(x)=\left(\frac{V(x)}{V^\prime(x)^2}\sqrt{\mu(x)}\right)^{\prime\prime}+
 \left(\frac{V(x)}{V^\prime(x)^2}\frac{d}{dx}\sqrt{\mu(x)}\right)^\prime$$
 where the primes denote differentiation with respect to  $x$. \\
 \textbf{Theorem 1}\;\;\;If $N(x)\ge 0$ for all $x\in K(E)$ the the period function $T$ is increasing on $(0, E)$. If
 $N(x)\le 0$ for all $x\in K(E)$ the the period function $T$ is decreasing on $(0, E)$.\\

 It is apparent that when $\mu(x)=1$  the function $N(x)$ reduces to $(V(x)/V^{\prime}(x)^2)^{\prime\prime}$ and the monotonicity of the period function therefore arises as a consequence of the convexity of the function $(V(x)/V^{\prime}(x)^2)$.\\

 Our motivation for inclusion of the mass function in the Hamiltonian stems from our earlier work on the isochronous behaviour of the Li\'{e}nard equation of the second kind \cite{CGG,GCG}. In fact several  equations of the Painlev\'{e}-Gambier classification belong to this family. Moreover the Jacobi Last Multiplier (JLM), $\mu(x)$, for Li\'{e}nard equations of the  second kind have the generic form $\mu(x)=\exp(2F(x))$ where $F(x)=\int ^x f(s) ds$ and plays the role of a mass term in the corresponding Lagrangian. Indeed by using the JLM it is possible to construct suitable transformations which map the Hamiltonian to that of a linear harmonic oscillator or to the isotonic oscillator which are known to be isochronous.\\

We illustrate our construction via examples. Rothe \cite{Rothe} defined some sets of functions appearing in the classical
system $ \dot{x} = -y, \dot{y} = f(x)$ giving rise to a monotonic energy-period function. 
He showed that the function $ f_1 = 5{f^{\prime \prime}}^2 - 3f^{\prime}f^{\prime \prime \prime}$ is one of the function
$f$ in this list. The monotonicity condition $f_1 > 0$ is called Schaaf's criteria \cite{Sr1,Sr2,Chouikha}. 
In this paper we find the position dependent
generalization of the $f_1$ function and Schaaf's criteria.
 The article is organized as follows. In section 2 we provide a proof of Theorem 1 and in section 3 we make a few observations regarding isochronicity. The content of this paper can be viewed as a contribution to the generalization of Chicone's result \cite{C1}.

 \section {Proof of Theorem 1}
 The period function for (\ref{L3}) is given by
 \be\label{T1}T=\sqrt{2}\int_{x_1^*}^{x_2^*} \frac{\sqrt{\mu(x)}dx}{\sqrt{E-V(x)}}.\ee
 Let $\xi=\int \sqrt{\mu(x)} dx=\eta(x)$ where it is assumed $\eta(x)$ is invertible. Then the period function may be written as
 $$T=\sqrt{2}\int_{\xi_1}^{\xi_2} \frac {d\xi}{\sqrt{E-V(\eta^{-1}(\xi))}}$$
Assuming $V(x)$ to have a quadratic minimum at the origin we set $r^2=V(\eta^{-1}(\xi))$ whence the period function becomes
$$T=\sqrt{2}\int_{-\sqrt{E}}^{\sqrt{E}} \frac{dr}{\sqrt{E-r^2}}\frac{1}{\left(\frac{V^\prime(\eta^{-1}(\xi))}{2\sqrt{V(\eta^{-1}(\xi))}}
\frac{d}{d\xi}(\eta^{-1}(\xi))\right)}$$
Using the substitution $r=\sqrt{E}\sin\theta$ and noting that $d(\eta^{-1}(\xi))/d\xi=dx/d\xi=1/\sqrt{\mu(\eta^{-1}(\xi))}$ we may rewrite the above integral as
$$T=\sqrt{2}\int_{-\pi/2}^{\pi/2} \frac{\sqrt{\mu(\eta^{-1}(\xi))}d\theta}{
\left(\frac{V^\prime(\eta^{-1}(\xi))}{2\sqrt{V(\eta^{-1}(\xi))}}\right)}$$
Let us further write $r=\sqrt{V(x)}=h(x)$ so that $T$ is expressible in the form
\be T=\sqrt{2}\int_{-\pi/2}^{\pi/2}\frac{\sqrt{\mu(h^{-1}(\sqrt{E}\sin\theta))}}
{h^\prime(h^{-1}(\sqrt{E}\sin\theta))}d\theta\ee
Taking the derivative of the period function with respect to the energy $E$ yields
\be\label{T5} \frac{dT}{dE}=-\frac{1}{\sqrt{2E}}\int_{-\pi/2}^{\pi/2}S(u)\sin\theta  d\theta\ee
where $$S(u)=\frac{h^{\prime\prime}(u)\mu(u)-\frac{1}{2}\mu^\prime(u)h^\prime(u)}{(h^\prime(u))^3 \sqrt{\mu(u)}}\;\;\;\mbox{with}\;\;\;u=h^{-1}(\sqrt{E}\sin\theta).$$
Using integration by parts the integral in (\ref{T5}) may be written as
\be\label{T6}\frac{dT}{dE}=\frac{1}{\sqrt{2}}\int_{-\pi/2}^{\pi/2}G(u) \cos^2\theta d\theta\ee where the function $G(u)=-S^\prime(u)/h^\prime (u)$. Carrying out the differentiation and bearing in mind that $h^2(x)=V(x)$ one can express the function $G(x)$ in the following form
$$G(x)=\frac{1}{8V^2 h^\prime(x)^5}\left[A(x)\sqrt{\mu(x)}-B(x)\frac{d}{dx}\sqrt{\mu(x)}+C(x) \frac{d^2}{dx^2}\sqrt{\mu(x)}\right]$$ where
$$A(x)=(6VV^{\prime\prime 2}-3V^{\prime 2}V^{\prime\prime}-2VV^\prime V^{\prime\prime\prime}),
B(x)=(6V V^\prime V^{\prime\prime}-3 V^{\prime 3}),
C(x)=2V V^{\prime 2}$$
It is obvious now that if the quantity within the square brackets denoted by say $N(x)$ is $> 0$ then the period function is monotonic increasing and if $N(x)< 0$ then the period function is decreasing. Indeed the expressions for $A(x), B(x)$ and $C(x)$ can be simplified further and we can express $G(x)$ is the succinct form
$$G(x)=\frac{2}{h^\prime(x)}N(x)=\frac{2}{h^\prime(x)}\left[\left(\frac{V}{V^{\prime 2}}\sqrt{\mu(x)}\right)^{\prime\prime}+\left(\frac{V}{V^{\prime 2}}\frac{d}{dx}\sqrt{\mu(x)}\right)^\prime\right]$$

 This completes the proof of Theorem 1.\\

\noindent
 Let $ u(x)=V(x)/V^{\prime 2}$. The function $N(x)$ may be expressed as
 \be\label{M1}N(x)=\left(\frac{V}{V^{\prime 2}}\sqrt{\mu(x)}\right)^{\prime\prime}+\left(\frac{V}{V^{\prime 2}}\frac{d}{dx}\sqrt{\mu(x)}\right)^\prime=u^{\prime\prime}\sqrt{\mu}+3u^\prime (\sqrt{\mu})^\prime +u(\sqrt{\mu})^{\prime\prime}.\ee
Clearly when $\mu=1$ the condition for monotonicity reduces to $u^{\prime\prime}>0(<0)$ which means that $u$ is concave up (concave down). 

\subsection{Illustration}

Following example is considered in \cite{KM}.

$$\ddot{x}-\frac{2px}{1+x^2}+g(x)=0$$
Here $f(x)=-2px/(1+x^2)$ and
$$F(x)=-\int_0^x\frac{2ps}{1+s^2}ds=\ln(1+x^2)^{-p}$$ as a result of which the mass function
$$\mu(x)=e^{2F(x)}=\frac{1}{(1+x^2)^p}>0 \;\;\forall x\in (-\infty, +\infty).$$
It is easy to deduce that
$$\frac{d}{dx}\sqrt{\mu}=\sqrt{\mu} f(x),$$
$$\frac{d^2}{dx^2}\mu=\sqrt{\mu}(f^2+f^\prime),$$ and as a consequence we may express $N(x)$ in the form
$$N(x)=\sqrt{\mu}\left[u^{\prime\prime}+3u^\prime f +2u(f^2+f^\prime)\right]$$ where $u=V/V^{\prime 2}$.

As $\sqrt{\mu}>0\;\;\forall x\in (-\infty, +\infty)$ therefore monotonicity of the period function reduces to the condition
$$M(x)=u^{\prime\prime}+3u^\prime f +2u(f^2+f^\prime)>0\;\; or\;\;(<0).$$
Here $$f^2+f^\prime=\frac{2p}{(1+x^2)^2}[(2p+1)x^2-1].$$
Consider the case when $p=(2m+1)/2$  with $m$ being a positive integer and recall that the potential function
$V(x)$ is given by
$$V(x)=\int \mu(x)g(x)dx+const.$$ where we assume
$$g(x)=x+a_3x^3+a_5x^5+\cdots +a_{2p}x^{2p}=\sum_{i=0}^m a_{2i+1}x^{2i+1}, \;\;a_1=1$$ Inserting the expression for $\mu(x)$ and assuming $V(0)=0$ it follows that
$$V(x)=\frac{1}{2}\sum_{i=0}^m\sum_{j=0}^i a_{2i+1}\left(\begin{array}{cc} i\\j\end{array}\right)\frac{(-1)^{i-j}}{2m-j}\left(1-\frac{1}{(1+x^2)^{2m-j}}\right),$$ and
$$V^\prime(x)=x\sum_{i=0}^m\sum_{j=0}^i a_{2i+1}\left(\begin{array}{cc} i\\j\end{array}\right)\frac{(-1)^{i-j}}{(1+x^2)^{2m+1-j}}$$
We now investigate a particular case  when $m=1$, i.e., $p=3/2$ for which the expressions for $V(x)$ and $V^\prime(x)$ are as follows
$$V(x)=\frac{x^2}{2(1+x^2)^2}\left(1+\frac{1+a_3}{2}x^2\right),\;\;\;V^\prime(x)=\frac{x}{(1+x^2)^3}(1+a_3x^2).$$ Hence the function
$$u(x)=\frac{V}{V^{\prime 2}}=\frac{1}{2}\frac{(1+x^2)^4}{(1+a_3x^2)^2}\left(1+\frac{1+a_3}{2} x^2\right).$$ It is clear that $u(x)>0$ for all $x\in (-\infty, +\infty)$ with $u(0)=1/2$. Furthermore
$u^\prime(x)=xu(x)K(x)$ where the function $K(x)$ is given by
$$K(x)=\left[\frac{8}{1+x^2}-\frac{4a_3}{1+a_3x^2}+\frac{1+a_3}{1+\frac{1+a_3}{2}x^2}\right],$$
and
$$u^{\prime\prime}(x)=u(x)\left[K(x)+x^2K^2(x)+xK^\prime(x)\right].$$
As  $u^\prime$ and $u^{\prime\prime}$ are both proportional to $u(x)$, which is positive if $a_3 > -1$, the monotonicity condition simplifies to the requirement
$$\frac{M(x)}{u(x)}=K(x)(1+3xf)+x^2K^2(x)+xK^\prime(x)+2(f^2+f^\prime)>0\;\; or (<0).$$

For $m=1$ i.e., $p=3/2$  we have $f(x)=-3x/(1+x^2) $ and $f^2+f^\prime =3(4x^2-1)/(1+x^2)^2$.
Using these expressions  and the form of $K(x)$ as stated above we obtain  upon simplification

$$C_{5}x^{10}+C_{4}x^{8}+C_{3}x^{6}+C_{2}x^{4}+C_{1}x^{2}+C_{0}>0\;\;\;or\;\;\;(<0)$$
as the condition for monotonicity. It is found that the coefficients
$$C_5=C_4=0, \;\;C_3=\frac{3}{4}(a_3+1)(a_3-1)(a_3^2+2a_3-4)$$
$$C_2=\frac{3}{2}(a_3-1)(3a_3^2+a_3-7),\;\;C_1=\frac{1}{4}(a_3-1)(21a_3-39),\;\;C_0=\frac{3}{2}(1-a_3)$$




\bigskip

\begin{figure}[h]
\centering
\begin{subfigure}{.5\textwidth}
  \centering
 \includegraphics[width=.7\linewidth]{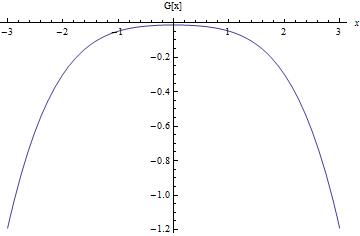}
 \caption{Graph for $a_3 = 1.001$}
 \label{fig:sub1}
\end{subfigure}%
\begin{subfigure}{.5\textwidth}
 \centering
\includegraphics[width=.7\linewidth]{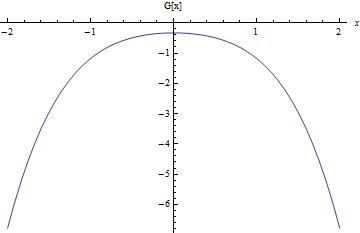}
\caption{Graph for $a_3 = 1.055$ }
\label{fig:sub2}
\end{subfigure}
\caption{Two subfigures $a_3>1$ shows monotonically decreasing}
\label{fig:test}
\end{figure}

\bigskip

\begin{figure}[h]
\centering
\begin{subfigure}{.5\textwidth}
 \centering
  \includegraphics[width=.7\linewidth]{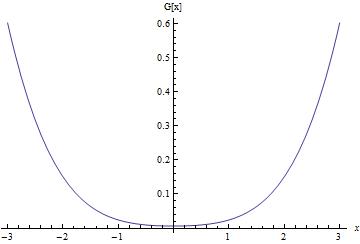}
 \caption{Graph for $a_3 = 0.999$}
 \label{fig:sub1}
\end{subfigure}%
\begin{subfigure}{.5\textwidth}
 \centering
  \includegraphics[width=.7\linewidth]{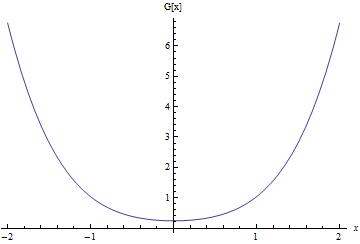}
 \caption{Graph for $a_3 = 0.96$ }
\label{fig:sub2}
\end{subfigure}
\caption{Two subfigures $a_3 < 1$ shows monotonically increasing }
\label{fig:test}
\end{figure}


\bigskip

\paragraph{Remark :} It is clear that when $a_3 = 1$, all the coefficients $c_i = 0$ for all $i = 1, \cdots 5$. 
Hence the system becomes isochronous.

\section{Isochronicity}
From theorem 1 it is clear that the period function will be independent of the energy $E$  if $N(x)=0$. The latter therefore provides a necessary condition for the determination of isochronous potentials for the Li\'{e}nard equation of the second kind given by (\ref{L2}). The following example provides an illustration of the usefulness of theorem 1 for testing isochronicity.\\

\noindent
\textit{Example} $\dot{x}=f(x)y$, $\dot{y}=-U(x)$ where $U(x)=\int_0^x ds/f(s)$.\\
Here $f(x)$ is assumed to be an even function such that $f(x)>0$. The above first-order system of equations is equivalent to the ODE
$$\ddot{x}-\frac{f^\prime(x)}{f(x)}\dot{x}^2+f(x)U(x)=0,$$
and comparison with (\ref{L3})  shows that $\mu(x)=1/f^2(x)$ and $V^\prime(x)=U(x)/f(x)$. As $U(x)=\int_0^x ds/f(s)$ it follows that $dU(x)=dx/f(x)$ and $U(x)dU(x)=U(x)dx/f(x)=V^\prime(x)$ so that $V(x)=U^2(x)/2$. Having obtained the explicit form of $V(x)$ it is easy to verify that
$$\frac{V}{V^{\prime 2}}\sqrt{\mu(x)}=\frac{1}{2}f(x),\;\;\; \frac{V}{V^{\prime 2}}\frac{d}{dx}\sqrt{\mu(x)}=-\frac{1}{2}f^\prime(x).$$
Consequently
$$N(x)=\left(\frac{V}{V^{\prime 2}}\sqrt{\mu(x)}\right)^{\prime\prime}+
\left(\frac{V}{V^{\prime 2}}\frac{d}{dx}\sqrt{\mu(x)}\right)^\prime=0,$$
and hence the system of equations constitutes an planar isochronous system.

 Writing $y=(V/V^{\prime 2})\sqrt{\mu}$ the condition $N(x)=0$ becomes
 $$\frac{d^2}{dx^2}(y\sqrt{\mu})=0\;\;\;\Rightarrow \frac{d}{dx}(y\sqrt{\mu})=const.$$
From (\ref{L2}) it follows that $\mu(x)=exp(2F(x))$ where $F(x)=\int_0^x f(s) ds$ and $V^\prime=ge^{2F}$. It follows that
$$y=\frac{V}{V^{\prime 2}}e^F=Ce^{-F}\int e^F dx+ De^{-F},$$
where $C$ and $D$ are constants of integration. Solving for $V$ we have
$$V=V^{\prime 2}\left[Ce^{-2F}\int e^Fdx+De^{-2F}\right]$$
and upon differentiating with respect to $x$ and eliminating $V^\prime$ we find that
$$2D(g^\prime+gF^\prime)+C\left[ge^F+2(g^\prime+gF^\prime)\int e^F dx\right]=1.$$
 Here $C$ and $D$ are arbitrary constants of integration. Let $P(x)=g^\prime+gF^\prime$, then the last relation
 becomes
 \be\label{W1} 2DP(x)+C\left[ge^F +2P(x)\int e^F dx\right]=1\ee
Differentiation with respect to $x$ gives
\be\label{W2} 2DP^\prime(x)+C\left[3P(x)e^F+2P^\prime(x)\int e^F dx\right]=0.\ee
The condition for existence of non-trivial solution of the arbitrary constants $C$ and $D$ requires the Wronskian of
(\ref{W1}) and (\ref{W2}) to be non-vanishing and this yields the condition:
\be\label{W3} W(x)=3P^2(x)-gP^\prime(x)\ne 0.\ee The solutions for $D$ and $C$ are given by
\be\label{W4}2D=\frac{3P+2P^\prime e^{-F}\int e^F dx}{W(x)},\ee
\be\label{W5} C=\frac{P^\prime e^{-F}}{W(x)}.\ee
The condition $dC/dx=0$  implies
\be\label{W*}3P^2P^\prime F^\prime+5PP^{\prime 2}-3P^2P^{\prime\prime}=0\ee
On the other hand the condition $dD/dx=0$ implies
\be\label{W**}\left(3P^2P^\prime F^\prime+5PP^{\prime 2}-3P^2P^{\prime\prime}\right)\left(ge^F+2P\int e^F dx\right)=0,\ee from which we obtain once again the condition (\ref{W*}) provided $ge^F+2P\int e^F dx\ne0$. We have therefore the following proposition.
\begin{prop}\label{P1} The condition for isochronicity of the equation $\ddot{x}+f(x)\dot{x}^2+g(x)=0$ when expressed in terms of the functions $f$ and $g$ is given by
 $$3P^2P^\prime F^\prime+5PP^{\prime 2}-3P^2P^{\prime\prime}=0$$ where $P(x)=g^\prime+gF^\prime$ and $F^\prime=f$ provided (i) $3P^2(x)-gP^\prime(x)\ne 0$ and (ii) $ge^F+2P\int e^F dx\ne 0$. \end{prop}

As the above proposition shows $P(x)=1$ always leads to the fulfillment of the preceding condition and leads to the well known condition for isochronicity $g^\prime+fg=1$ obtained in \cite{S1, GCG,Chouikha2}

\smallskip

\paragraph{Conservative system}
For a conservative system the equation of motion becomes
\be \ddot{x}+V^\prime(x)=0 \ee or in other words $\mu(x)=1$.

In this case as $f=F^\prime=0$ and Proposition (\ref{P1}) reduces to
$$5PP^{\prime 2}-3P^2P^{\prime\prime}=0, $$ with $P(x)=g^\prime$. This leads to the condition
\be\label{W6}5g^\prime g^{\prime \prime 2} -3g^{\prime 2}g^{\prime\prime\prime}=0.\ee
Eqn. (\ref{W6}) corresponds to the isochronous counterpart of the condition derived by Schaaf in \cite{Sr2,Chouikha} 
for monotonicity, namely
\be\label{W7} 5g^\prime g^{\prime \prime 2} - 3g^{\prime 2}g^{\prime\prime\prime} \ge 0.\ee
The Schaaf's criteria states that the center of system $\dot{x} = -y, \dot{y} = V^{\prime}(x)$ has a monotonically
increasing period function in the case that satisfies (\ref{W7}). 

\section{Outlook}

 In this paper we have given the generalization of Chicone's condition for monotonicity for the case
of Hamiltonian planar vector field with position dependent mass. This situation automatically appears in
the Li\'enard II type equation. We have also derived the isochronicity criteria from the a null condition.
We illustrate our result with a suitable example.

\section*{Acknowledgments}
We would like to thank Professor Carmen Chicone and Professor Jean-Marie Strelcyn
for their help and valuable comments in the preparation
of the manuscript. We would also like to thank Mr. Ankan Pandey for computational assistance.

\end{document}